\documentclass{ws-procs9x6}

\begin{document}

\title{Polarized electron emission from strained G\lowercase{a}A\lowercase{s}/G\lowercase{a}A\lowercase{s}P superlattice photocathodes \footnote{\uppercase{T}his work was supported in part by the \uppercase{U.~S.~D}epartment of \uppercase{E}nergy under contract numbers \uppercase{DE-AC}02-76\uppercase{SF}00515 (\uppercase{SLAC}), and \uppercase{DE-AC}02-76\uppercase{ER}00881 (\uppercase{UW}).}}

\author{T. Maruyama$^\dagger$, D.-A. Luh$^\dagger$, A. Brachmann$^\dagger$,
J.~E. Clendenin$^\dagger$, E.~L. Garwin$^\dagger$,
S. Harvey$^\dagger$, J. Jiang$^\dagger$, R.~E. Kirby$^\dagger$, \\ 
C.~Y. Prescott$^\dagger$, 
R. Prepost$^\ddagger$ and A.~M. Moy$^*$}
\address{
   $^\dagger$Stanford Linear Accelerator Center,
   Menlo Park, California~94025 \\
   $^\ddagger$Department of Physics, University of Wisconsin,
   Madison, Wisconsin~53706 \\
   $^*$SVT Associates, Inc., Eden Prairie, Minnesota~55344}

\maketitle

\abstracts{
Spin-polarized electron photoemission has been studied for
GaAs/GaAs$_{1-x}$P$_x$ strained superlattice cathodes grown by gas-source
 molecular beam epitaxy.
The superlattice structural parameters are systematically varied
to optimize the photoemission characteristics.
The heavy-hole and light-hole transitions are reproducibly observed in
quantum efficiency spectra, enabling direct measurement of the band energies
and the energy splitting.
Electron-spin polarization as high as 86\% with over 1\% quantum efficiency
has been observed.}

\section{Introduction}
Single strained GaAs photocathodes were introduced for the SLAC polarized 
electron source in 1993. After 10 years of experience with many cathode 
samples, the maximum polarization using the GaAs/GaAsP single strained-layer 
cathode remains limited to ~80\%, while the quantum efficiency (QE) for 
a 100-nm epilayer is only 0.3\% or less. Two known factors limit the
polarization of these cathodes: 1) a limited band splitting; and
2) a relaxation of the strain in the surface epilayer
since the 10-nm critical thickness for the 1\% lattice-mismatch is exceeded.
Strained superlattice
structures, consisting of very thin quantum well layers alternating with
lattice-mismatched barrier layers are excellent candidates for achieving
higher polarization since they address these two issues.
Due to the difference in the effective mass of the heavy- and
light-holes, a superlattice exhibits a natural splitting of the valence band,
which adds to the strain-induced splitting. In addition, each of the
superlattice layers is thinner than the critical thickness. 
This paper presents an investigation
 of strained GaAs/GaAsP superlattice samples in which the principal structural
 parameters are systematically varied to define the optimum
structural details \cite{tm}.

\section{Experiment}
Four parameters specify the superlattice structure: the GaAs well width,
the GaAs$_{1-x}$P$_x$ barrier width, the phosphorus fraction, and
the number of periods. Table 1 summarizes the eleven superlattice samples
studied here.

\begin{table}[ph]
\tbl{Strained GaAs/GaAs$_{1-x}$P$_x$ superlattice samples.}
{\footnotesize
\begin{tabular}{@{}ccccc@{}}
\hline
{} &{} &{} &{} &{} \\[-1.5ex]
Sample & Well (nm) & Barrier (nm) & x & No. Period\\[1ex]
\hline
{} &{} &{} &{} &{}\\[-1.5ex]
  1   & 4 & 4 & 0.25 & 12 \\[1ex]
  2   & 4 & 4 & 0.30 & 12 \\[1ex]
  3   & 4 & 4 & 0.36 & 12 \\[1ex]
  4   & 4 & 4 & 0.40 & 12 \\[1ex]
  5   & 4 & 4 & 0.36 &  9 \\[1ex]
  6   & 4 & 4 & 0.36 & 15 \\[1ex]
  7   & 4 & 4 & 0.36 & 20 \\[1ex]
  8   & 4 & 4 & 0.36 & 30 \\[1ex]
  9   & 3 & 3 & 0.36 & 16 \\[1ex]
 10   & 4 & 3 & 0.36 & 14 \\[1ex]
 11   & 5 & 3 & 0.36 & 12 \\[1ex]
\hline
\end{tabular}\label{table2} }
\vspace*{-13pt}
\end{table}

Figure 1 shows the polarization and QE as a function of the excitation
photon energy for Sample 3.
The peak polarization is 86\% with 1.2\% QE. The QE spectrum shows two
distinct steps as expected from the density of states for
the two dimensional structure.
The first step corresponds to the heavy-hole (HH) band to the conduction band
excitation, while the second step corresponds to the light-hole (LH) band to
the conduction band excitation.
The energy splitting between the HH- and LH-bands is 82 meV.

\begin{figure}[ht]
\centerline{\epsfxsize=2.5in\epsfbox{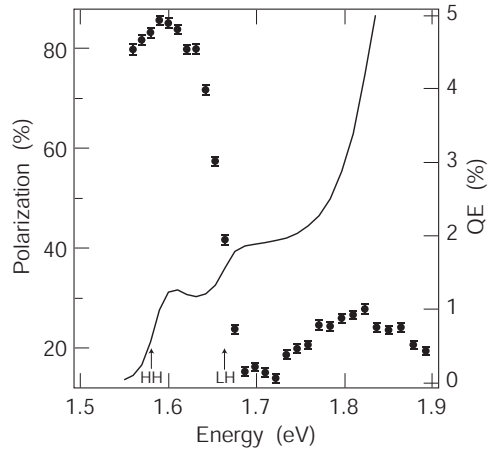}}
\caption{Polarization and QE as a function of excitation light energy}
\end{figure}

When the phosphorus fraction is varied, the lattice-mismatch between
the well and the barrier changes, thus the superlattice strain can be varied.
While a larger phosphorus fraction generates a larger strain and therefore
a larger energy splitting between the HH and LH bands, the strain
within a layer may relax.
For Samples 1, 2, 3, and 4, the phosphorus fraction was increased from
0.25 to 0.40 keeping the total superlattice thickness constant.
Figure 2 shows the peak polarization and QE anisotropy \cite{qe} as a function of
the phosphorus fraction for constant total thickness. The measured HH-LH energy
splitting is also shown in the figure.
Although the HH-LH energy splitting increased from 60 meV
to 89 meV, the peak polarization and the QE anisotropy did not change
significantly at about 85\% and 1.7\%, respectively,
indicating that this degree of  energy splitting is sufficient to maximize
the spin polarization.

\begin{figure}[ht]
\centerline{\epsfxsize=2.5in\epsfbox{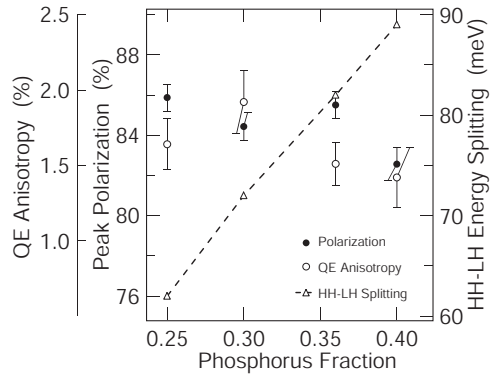}}
\caption{Peak polarization (solid circles), QE anisotropy (open circles), and measured HH-LH energy splitting (triangles) as a function of the phosphorus fraction.}
\end{figure}

Using samples with the same well (4 nm)/barrier (4 nm) thickness and
phosphorus fraction ($x$=0.36), the total superlattice thickness was varied.
Figure 3 shows the peak polarization and QE anisotropy as a function of the
number of superlattice periods using Samples 3, 5, 6, 7, and 8.
Also shown in Figure 3 is the strain relaxation in the superlattice GaAs well
layers measured using x-ray diffraction.
Although the well width is smaller than the critical thickness, increased
superlattice periods will result in strain relaxation.
As the strain relaxation steadily increases with the superlattice thickness,
the peak polarization and QE anisotropy appear constant at 85.5\%,
and 1.5\%, respectively, for less than 15 periods.
At more than 20 periods, however, the peak polarization decreases and the
QE anisotropy increases rapidly.

\begin{figure}[ht]
\centerline{\epsfxsize=3in\epsfbox{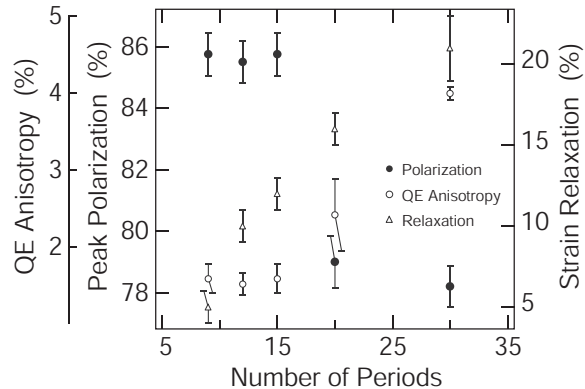}}
\caption{Peak polarization (solid circles), QE anisotropy (open circles), and strain relaxation (triangles) as a function of the superlattice period.}
\end{figure}

The QE and polarization spectra can be understood in terms of the superlattice
band structure.
In particular, the band structure is very sensitive to the well width.
Using samples with the same barrier thickness (3 nm) and phosphorus fraction
($x$=0.36), the well thickness was varied while the number of periods was
adjusted to keep the same total superlattice thickness.
Figure 4 shows the spin polarization as a function of the excitation
photon energy for Samples 9, 10, and 11.
As the well thickness was increased, the polarization spectra shifted towards
lower energy. The peak polarization was, however, independent of
the well thickness.
While the changes in the band structure differentially affected
the polarization spectra for energies above the polarization peak,
the maximum polarization remained constant at about 86\%,
indicating that the valence-band splitting for this range of well
thicknesses was sufficient.

\begin{figure}[ht]
\centerline{\epsfxsize=2.5in\epsfbox{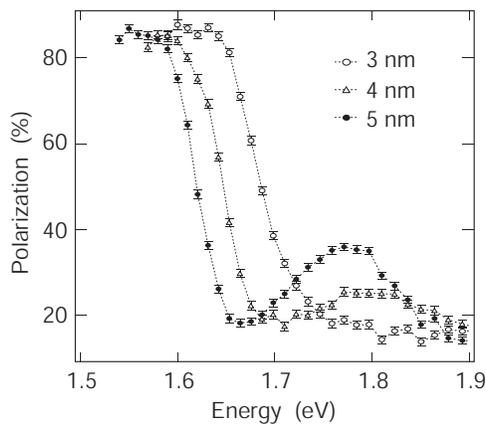}}
\caption{Polarization as a function of the excitation light energy for the three samples with a different well width; 3 nm (open circles), 4 nm (triangles), and 5 nm (solid circles).}
\end{figure}

\section{Conclusions}
We have investigated polarized photoemission from strained
GaAs/GaAsP superlattice structures by systematically varying the superlattice
parameters. The heavy- and light-hole excitations have been
observed for the first time in the QE spectra, enabling direct measurements
of the heavy- and light-hole energy bands.
Spin polarization as high as 86\% is reproducibly observed with the QE
exceeding 1\%. The superlattice structures have superior
polarization and QE compared to the single strained-layer structures of
GaAs/GaAsP photocathodes.

\end{document}